\documentclass[prb,floats,twocolumn]{revtex4-2}
\usepackage[english]{babel}
\usepackage{xcolor}
\usepackage{amsmath}
\usepackage{graphicx}
\usepackage{epstopdf}
\newcommand{\bk}{{\bf k}}

\graphicspath{{figures/}}

\begin{document}
\title{Mechanism for Switchability in Electron-Doped Ferroelectric Interfaces}
\date{\today}
\author{Kelsey S. Chapman}  \email{kelseychapman@trentu.ca}
\author{W. A. Atkinson} \email{billatkinson@trentu.ca}
\affiliation{Department of Physics and Astronomy, Trent University, Peterborough, Ontario K9L 0G2, Canada}

\begin{abstract}
With the recent experimental verification that ferroelectric lattice distortions survive in the metallic phase of some materials, there is a desire to create devices that are both switchable and take advantage of the novel functionalities afforded by polar interfaces. In this work, we explore a simple model for such an interface and demonstrate a mechanism by which a metallic ferroelectric substrate may be switched by a bias voltage. This finding is in contrast to the reasonable expectation that hysteresis is prevented by screening of external fields in ferroelectric metals. Instead, the electron gas binds to polarization gradients to form a compensated state. Uncompensated electrons, which  may screen external fields, are generated either when the electron density exceeds the ferroelectric polarization or when the bias voltage exceeds a ``spillover'' threshold.  We propose that switchable thin films may be optimized by choosing an electron density that is slightly less than the lattice polarization.  In addition to the high-polarization states, we find that thin metallic films also have a low-polarization state with average polarization near zero.  
Unlike in insulating films, where the polarization is small everywhere in this state, the low-polarization state in the metallic films consists of two head-to-head domains of opposite polarization. This domain formation is enabled by the screening of depolarizing fields by the electron gas.
\end{abstract}
\maketitle

\section{Introduction}

The concept of  ferroelectric (FE) metals dates back to Anderson and Blount,\cite{anderson_symmetry_1965} who in 1965 argued that second-order phase transitions observed in some metals might be an indication of a FE-like structural transition.   However, there was until recently a general belief that itinerant electrons destroy polar lattice distortions because the latter is driven by attractive Coulomb forces between ions, which are screened in metallic systems.\cite{Lines:2001bn}  Metallicity is certainly antagonistic towards ferroelectricity, but it is now understood that FE-like lattice polarizations persist in some metals. This persistence points to a role for short-range interactions, in addition to long-range Coulomb forces, in stabilizing polar lattice distortions \cite{cohen_origin_1992,posternak_role_1994,zhao_meta-screening_2018},  
and the relationship between electron (or hole) doping and polar lattice distortions remains an active area of research \cite{puggioni:2014,benedek_ferroelectric_2016,zhao_meta-screening_2018,fu_possible_2020,michel:2021}.

In practice,  while there are a few known stoichiometric metals with an intrinsically FE-like phase transition,\cite{shi_ferroelectric-like_2013,fei_ferroelectric_2018,sharma_room-temperature_2019,kim_polar_2016} most of the attention has focused on carrier-doping known FE perovskites to achieve metallicity.  In transition metal perovskites, such as Sr$_{1-x}$Ba$_x$TiO$_3$ or Sr$_{1-x}$Ca$_x$TiO$_3$, doping may be achieved by oxygen depletion or cation substitution  \cite{kolodiazhnyi_persistence_2010,cordero_probing_2019,fujioka_ferroelectric-like_2015,takahashi_polar_2017,Rischau:2017vj,engelmayer_ferroelectric_2019}.

Alternatively, metallicity may be achieved by an ``electronic reconstruction'' that occurs in certain heterostructures and bilayers.  This mechanism   was first demonstrated by Ohtomo {\em et al.} for the non-FE SrTiO$_3$ interfaces  LiTiO$_3$/SrTiO$_3$ and LaAlO$_3$/SrTiO$_3$ \cite{Ohtomo:2002ib,ohtomo04}.  In these systems, the cap layer (LiTiO$_3$ or LaAlO$_3$) has a polar unit cell while the substrate (SrTiO$_3$) is non-polar.  The polarization is thus discontinuous at both the cap surface and the interface, and this generates internal electric fields that tend to transfer electrons from the surface to the interface.  Because SrTiO$_3$ has a narrower band gap than either cap layer, the resulting two-dimensional electron gas (2DEG) lies on the SrTiO$_3$ side of the interface (see Ref.~\cite{Gariglio:2015jx} for a review).  These systems are especially attractive because the electron density, and associated electronic  properties, can be tuned via back- and top-gating.\cite{Caviglia:2008uh,Caviglia:2010jv,Bert:2012el,eerkes_modulation_2013,Joshua:2013wl,Smink:2017cc,Raslan:2018}  

The extension to interfaces involving FE materials is natural, and there have been a number of theoretical proposals based on density functional theory for combinations of materials that can generate self-doped interfaces.\cite{Niranjan:2009ix,wang_first-principles_2009,AguadoPuente:2015if,yin_two-dimensional_2015,Fredrickson:2015cz,nazir_towards_2015} 
There are comparatively few experimental realizations, but significant steps forward have been taken in the past few years.  Zhou {\em et al.} \cite{zhou_artificial_2019} reported the coexistence of a 2DEG with a FE-like lattice polarization in the Sr$_{0.8}$Ba$_{0.2}$TiO$_3$ layer of an LaAlO$_3$/Sr$_{0.8}$Ba$_{0.2}$TiO$_3$ interface. More recently, Refs.~\cite{brehin_switchable_2020} and \cite{Tuvia2020}  investigated ferroelectricity and conductivity in Ca-doped SrTiO$_3$ interfaces. Br\'ehin {\em et al}.\ demonstrated that a switchable 2DEG forms in the perovskite layer of an Al/Sr$_{0.99}$Ca$_{0.01}$TiO$_3$ bilayer \cite{brehin_switchable_2020}, while Tuvia {\em et al.}\ observed hysteretic polarization and resistivity in Ref.~\cite{Tuvia2020} at a simultaneously ferroelectric and superconducting LaAlO$_3$/Sr$_{1-x}$Ca$_x$TiO$_3$ interface.  In all three systems, the perovskite substrate is intrinsically insulating, and charge carriers are provided by the cap layer.

That one could obtain a switchable polarization is not obvious, and indeed the conventional line of thought asserts that in FE metals external electric fields must be screened by the electron gas, making control of the polarization state impossible. For this reason, FE metals are often termed ``polar metals'' to emphasize the lack of switchability.  This raises an important question:  To what extent are external fields actually screened?  To answer this, we need to understand how, in fact, the polarization and electron gas are distributed throughout the FE substrate, and how this changes as a function of bias voltage.  Perhaps most importantly, we want to find a mechanism by which the direction of the lattice polarization may be switched by an external field.  

To address these questions, we model an interface between an insulating cap layer and an intrinsically insulating FE.   As with the conventional LaAlO$_3$/SrTiO$_3$ interfaces, the FE is presumed to become metallic as a result of charge transfer from the surface of the cap layer.  We are particularly motivated by systems in which a quantum paraelectric, such as SrTiO$_3$ or KTaO$_3$, is made FE, possibly by cation substitution or by strain.  Although their transition temperatures $T_c$ are typically quite low, SrTiO$_3$- or KTaO$_3$-based FEs are attractive because of their tunability and because of the prospect that superconductivity survives in the FE state.\cite{Rischau:2017vj,liu:2021,chen:2021a,chen:2021}  Furthermore, typical two-dimensional (2D) charge  densities for the electron gas are $en_\mathrm{2D}$
$\sim 5$~$\mu$C/cm$^2$,\cite{Gariglio:2015jx} which is comparable to observed polarizations in SrTiO$_3$-based FEs.\cite{menoret:2002,deLima:2015te} As a result, the 2DEG and lattice polarization make similar contributions to the internal electric field, and (as we show below) feedback between the two has a profound effect on their structures.

We note that there is an important distinction between the kinds of interfaces modeled in this work and so-called polar or ferroelectric metals.  For the latter, one generally has in mind materials in which the electron gas is uniformly distributed throughout the sample when the temperature is above the Curie temperature. For the interfaces studied here, however, the charge reservoir is presumed to be the surface of the cap layer, and above $T_c$ the electron gas is bound to within a few nm of the interface by the positive residual charge on the cap layer. 

Our calculations find self-consistent solutions for the polarization that are both switchable and show hysteresis as a function of bias voltage.  We are further able to identify key length scales that determine the electronic and polarization profiles within the FE substrate.  Our main result is that, at low electron densities, the 2DEG forms a cooperative state with the lattice polarization, in which charges associated with polarization gradients are compensated by the electron gas.  The effect of this is to screen depolarizing fields but leave external fields unscreened.  At high electron densities, the 2DEG develops a second component that is not bound to the polarization and which screens external fields.  The screening is partial, so hysteresis persists.

The model on which our calculations are based is outlined in Sec.~\ref{sec:method}.  In Sec.~\ref{sec:results}, we show hysteresis curves and present a detailed discussion of the polarization and electron gas profiles for different states along the hysteresis curves.  A brief discussion follows in Sec.~\ref{sec:discussion}.

\section{Method}
\label{sec:method}

\begin{figure}
\includegraphics[width=\columnwidth]{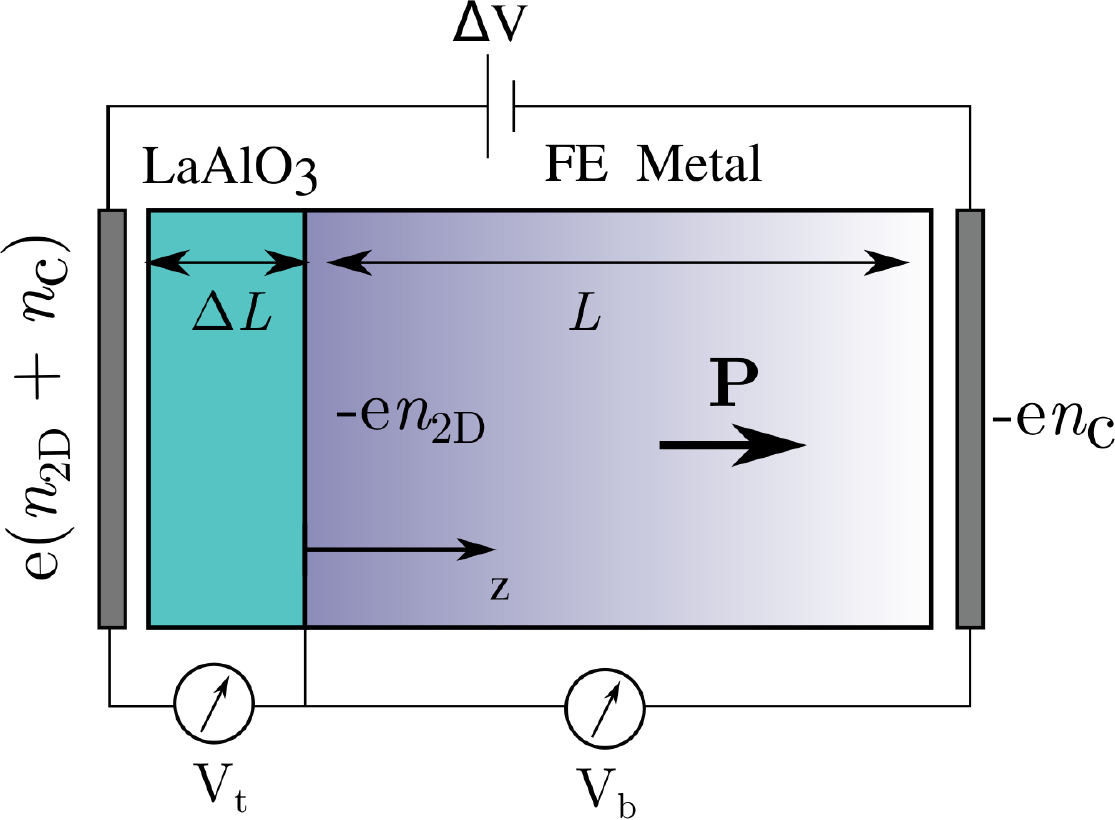}
\caption{Schematic of the model  interface.  The interface between a polar cap layer, for example LaAlO$_3$, and an intrinsically non-polar substrate, for example SrTiO$_3$, creates a 2DEG with a fixed electron density $n_\mathrm{2D}$.  A dilute concentration of Ca or Ba ions renders the substrate  ferroelectric. The profile of the lattice polarization $P$ and electron gas density may be controlled by an applied voltage $\Delta V$ across the device, which transfers charge $en_c$ between the electrodes on the sample surfaces while keeping $n_\mathrm{2D}$ fixed.   For convenience, we write $\Delta V = V_t + V_b$, where $V_t$ and $V_b$ are the voltages across the top cap layer and the bottom substrate, respectively.}
\label{fig:schematic}
\end{figure}

We adopt the geometry in Fig.~\ref{fig:schematic}, which shows an interface between a polar cap layer (for example, LaAlO$_3$) and a FE substrate (for example, Sr$_{1-x}$Ca$_x$TiO$_3$). 
The polarity of the cap layer comes from alternating layers of (LaO)$^{1+}$ and (AlO$_2$)$^{1-}$; these produce internal electric fields that generate a potential difference between the surface and the interface.\cite{Gariglio:2015jx} In non-FE LaAlO$_3$/SrTiO$_3$ interfaces, this potential difference is compensated by, among other things,\cite{Pentcheva:2009ef} a transfer of electrons from the surface (likely from defect states\cite{Bristowe:2014fc,lemal_polarity-field_2020}) to the interface. The resultant 2DEG is bound to the interface by the residual positive charge on the LaAlO$_3$ surface, and it can be manipulated by the application of a bias voltage across the sample.  Here, we presume that the FE substrate is electron-doped by a similar mechanism.

Figure~\ref{fig:schematic} defines a number of important model parameters.  The cap and substrate layers have thicknesses $\Delta L$ and $L$, respectively, and the substrate comprises $N_L$ one-unit-cell-thick monolayers, such that $L= N_L a$, with $a$ the substrate lattice constant. We treat the monolayers as discrete elements labeled by index $i_z$, measured from the interface. The total 2D electron density in the substrate is then
\begin{equation}
	n_\mathrm{2D} = \int_0^L n(z) dz = \frac{1}{a^2} \sum_{i_z=0}^{N_L-1} n_{i_z},
\end{equation}
where $n(z)$ is the 3D electron density and $n_{i_z}$ is the number of conduction electrons per unit cell in layer $i_z$. The electron gas is presumed to be manipulated by a bias voltage $\Delta V$ across capacitor plates on the top and bottom surfaces of the system.  The net charge on the surface of the cap layer then has a contribution $en_\mathrm{2D}$ from the charge transfer to the interface, and a second contribution $e n_c$ from the capacitor plates.   $\Delta V$ can be split into two contributions:  the voltage drop $V_t$ across the (top) cap layer, and the drop $V_b$ across the (bottom) substrate.    

Any two of the parameters $(n_\mathrm{2D}, \Delta V, V_t, V_b)$  can be chosen as independent, with the other two being dependent.  To keep doping-related effects distinct from charge-redistribution effects, we take $n_\mathrm{2D}$ as an independent variable.  It is then conceptually cleanest to take the second independent variable to be $\Delta V$, as it transfers charge between the capacitor plates without changing $n_\mathrm{2D}$.  In calculations, one can equally choose $n_\mathrm{2D}$ and $V_b$ as independent since $n_\mathrm{2D}$ is fixed by the chemical potential, although in experiments this  introduces the complication that $V_t$ must be fine-tuned for each $V_b$ to obtain a fixed value of $n_\mathrm{2D}$.  Nonetheless, we find it convenient to report results in terms of $V_b$  because it is more directly connected to the substrate polarization than $\Delta V$, and because it is not affected by potentials arising from crystal fields in the polar cap layer. These fields generate an offset that should be added to both $V_t$ and $\Delta V$, but that has no effect on $V_b$.

In the classical limit, electric fields in the substrate are fully screened by the 2DEG such that $V_b=0$ and $\Delta V = V_t$; in this limit the substrate polarization cannot be influenced by $\Delta V$ and, crucially, is not switchable by an applied voltage. As we show below, quantum effects reduce the screening in our model substrate sufficiently that the polarization can be manipulated by the bias voltage, and is switchable.

The polarization $P(z)$ is presumed to be perpendicular to the interface, along the $z$-axis, and to be uniform in the $x$ and $y$ directions.  In general, we cannot rule out the possibilities that other polarization axes might be relevant, or that domains might form such that $P$ is also a function of $x$ or $y$.  Indeed, electrostatic considerations for insulating FE films show that depolarizing fields may be reduced  by orienting the polarization parallel to the largest surfaces (i.e.\ perpendicular to the $z$ axis) \cite{Littlewood_review_2017}, or by forming Kittel domains  \cite{Kittel:1946,lukyanchuk:2018}.  We assume that the polarization direction is pinned to the $z$-axis by a combination of interfacial effects and external fields due to the cap layer, as it is in non-FE LaAlO$_3$/SrTiO$_3$ interfaces \cite{Gariglio:2015jx}.   This assumption  is supported by the observation of a large perpendicular polarization component near Sr$_{1-x}$Ca$_x$TiO$_3$ interfaces \cite{brehin_switchable_2020,Tuvia2020}.  Furthermore, we believe that Kittel domains will be suppressed by the internal screening of depolarizing fields by the electron gas.  We note that there is some experimental support for this assumption \cite{Tuvia2020}, although it remains to be tested by a future calculation. 

Our model comprises an electronic Hamiltonian, describing the 2DEG, and an ionic Hamiltonian, describing the lattice polarization. The lattice and electronic degrees of freedom are coupled through the self-consistently determined electric field. This model, described in detail below, was developed in Ref.~\cite{chapman_modified_2020} where it was applied to LaAlO$_3$/SrTiO$_3$ interfaces.

For the interface geometry pictured in Fig.~\ref{fig:schematic}, the electronic Hamiltonian takes the form
\begin{multline}
\hat H_\mathrm{el} = \sum_{i_z, \bk} \sum_{ \alpha, \sigma}
\big [ \epsilon_{i_z \bk \alpha} c^\dagger_{i_z\bk \alpha\sigma} c_{i_z\bk\alpha\sigma} \\
 + t_{\alpha, z} ( c^\dagger_{i_z+1\, \bk \alpha \sigma} c_{i_z\bk \alpha \sigma} +  c^\dagger_{i_z \bk \alpha \sigma} c_{i_z+1\, \bk \alpha \sigma} ) \big ].
 \label{eq:Hel}
\end{multline}
Here $\bk$ is a 2D wavevector describing translational degrees of freedom parallel to the interface; $\sigma$ labels the electron spin, and the index $\alpha$ labels the orbital degrees of freedom within each unit cell.  For FEs based on doped SrTiO$_3$, the relevant bands are derived from three Ti $t_{2g}$ orbitals ($\alpha = xy,\, yz,\, xz$).  The operator $c^\dagger_{i_z\bk \alpha \sigma}$ thus creates an electron in a 2D Bloch state in layer $i_z$ with orbital character $\alpha$ and spin $\sigma$.  

The electronic structure of the substrate is determined by a small number of tight-binding parameters.  For SrTiO$_3$, the nearest-neighbor Ti-Ti hopping matrix element $t_{\alpha, \delta}$ depends on both the orbital symmetry and the direction of hopping $\delta$, with 
\begin{equation}
t_{\alpha,\delta} = \left \{ \begin{array}{ll} 
t_{\perp}, & \mbox{ $\delta$ is perpendicular to the plane of $\alpha$}\\
t_{\|}, & \mbox{ $\delta$ lies in the plane of $\alpha$}
\end{array}\right ..
\end{equation}
Thus, interlayer hopping between $d_{xy}$ orbitals has the matrix element $t_{xy, z} = t_\perp$.  The layer-dependent dispersion in Eq.~(\ref{eq:Hel}) is
\begin{eqnarray}
\epsilon_{i_z \bk \alpha}&=& -2t_{\alpha, x} \cos(k_x a) - 2t_{\alpha, y} \cos(k_y a) - eV_{i_z}, \\
&\approx& -2(t_{\alpha, x}+t_{\alpha, y}) + \frac 12 \left ( \frac{k_x^2}{m_{\alpha, xx}}
+\frac{k_y^2}{m_{\alpha, yy}} \right ) - eV_{i_z},
\end{eqnarray}
where $V_{i_z}$ is the electrostatic potential in layer $i_z$, $-e$ is the electron charge, and $m_{\alpha, \delta \delta}^{-1} = 2t_{\alpha, \delta} a^2$, where $a=3.9$~\AA{} is the lattice constant of SrTiO$_3$.  The second form is used in this work because the band fillings are low.

The total electron density in layer $i_z$ is given by
\begin{equation} \label{eq:niz}
n_{i_z} = \sum_{\alpha, n} \eta_{\alpha, n} |\Psi_{\alpha, n}(i_z)|^2,
\end{equation}
with the eigenstates of the electronic Hamiltonian $\Psi_{\alpha, n}(i_z)$. (Note that the Hamiltonian does not mix orbital types, so the eigenstates are independent of $\bk$.) $\eta_{\alpha, n} = N_k^{-1} \sum_\bk f(\epsilon_{\alpha, n\bk})$ is the band filling for orbital type $\alpha$ and band $n$, with $N_k$ total $k_x$- and $k_y$-points and the Fermi-Dirac distribution $f(\epsilon_{\alpha, n\bk})$.

The lattice polarization may be described by a number of phenomenological approaches, including Landau-Ginzburg-Devonshire (LGD) theory, nonlinear phonon Hamiltonians \cite{Schneider:1976vn,Atkinson:2017jt}, and  pseudospin models \cite{Prosandeev:1999tx,chapman_modified_2020}.  At the mean-field level, all three approaches yield similar results \cite{Prosandeev:1999tx},  and reasonable quantitative fits to e.g.\ dielectric susceptibility measurements are possible by fitting model parameters.  Additional insights are possible depending on the model:  Phonon Hamiltonians highlight connections to quantum critical phenomena \cite{Atkinson:2017jt} but are computationally intensive; pseudospin  approaches are easily adapted to certain situations (e.g.\ modeling alloys \cite{kleemann00}) and require similar computational effort to LGD models.  Here, we use a pseudospin phenomenology and refer the reader to Ref.~\cite{chapman_modified_2020} for an extensive discussion of its validity as a model of SrTiO$_3$-based heterostructures.

The lattice polarization is modeled with a modified transverse Ising model,\cite{chapman_modified_2020} which for the layered geometry takes the form
\begin{multline}
\hat{H}_\mathrm{TIM} = -\frac{\Omega}{S} \sum_{i_z} \hat{S}^{x}_{i_z} -\frac{J_1}{S^2} \sum_{i_z} \hat S^z_{i_z}  \hat S^z_{i_z+1} \\
- \frac{J_0 -  2J_1}{2S^2}  \sum_{i_z} \hat S^z_{i_z} \hat S^z_{i_z}  - \frac{\mu}{S} \sum_{i_z} E_{i_z} \hat S^z_{i_z},
\label{eq:HTIM0}
\end{multline}
where $\hat S^a_{i_z}$ are components of a magnitude-$S$ pseudospin operator ${\bf \hat S}_{i_z}$.  The $x$-, $y$-, and $z$-directions of the pseudospin space are unrelated to physical directions; rather, the $z$-component of the pseudospin vector determines $P_{i_z}$, the mean z-component of the polarization  in layer $i_z$, while the $x$- and $y$- components allow for quantum dynamics of the pseudospin.   The polarization is
\begin{equation} \label{P}
P_{i_z} = \frac{\mu}{v_\mathrm{u.c.}} \frac{S^{z}_{i_z}}{S},
\end{equation}
where 
\begin{equation}
S^z_{i_z} = \langle \hat S^z_{i_z} \rangle,
\end{equation}
$\mu$ is the value of the saturated dipole moment of a single unit cell, and $v_\mathrm{u.c.}$ is the unit cell volume.   Of the remaining parameters in Eq.~(\ref{eq:HTIM0}), $\Omega$ and $J_0$ determine the bulk transition temperature and low-$T$ polarization, while $J_1$ is proportional to the square of the dipole-dipole correlation length \cite{chapman_modified_2020}. $E_{i_z}$ is the electric field in layer $i_z$.  Long-range dipole interactions are included implicitly in the electric field through the solution of Gauss' law, below.

Within mean-field theory, 
\begin{equation}
\hat H_\mathrm{TIM} \approx - \sum_{i_z} {\bf h}_{i_z} \cdot {\bf \hat S}_{i_z},
\label{eq:HTIM}
\end{equation}
with a Weiss mean field
\begin{equation}
{\bf h}_{i_z} = \frac 1 S \left (\begin{array}{ccc} \Omega , 0,
\dfrac {J_1}{S} \nabla_z^2 S^z_{i_z}
+ \dfrac{J_0}{S}  S^z_{i_z}  + \mu E_{i_z}
\end{array} \right ),
\label{eq:h}
\end{equation}
and
\begin{eqnarray}
\nabla_z^2 S^z_{i_z} &\equiv& S^z_{i_z-1} - 2 S^z_{i_z} + S^z_{i_z+1}. \label{eq:D2}
\end{eqnarray}
In Eq.~(\ref{eq:D2}), $S^z_{i_z\pm 1}$ are set to zero if $i_z\pm 1$ is outside of the system.  Self-consistency of Eqs.~(\ref{eq:HTIM}) and (\ref{eq:h}) requires that $S^z_{i_z}$ satisfy the mean-field equation
\begin{equation}
S^z_{i_z} =  S \frac{h^z_{i_z}}{|{\bf h}_{i_z}|} B_S\left (\frac{|{\bf h}_{i_z}|}{k_BT} \right ),
\label{eq:Bfn}
\end{equation}
where $k_B$ is the Boltzmann constant, $T$ is temperature, and $B_S()$ is the Brillouin function for pseudospin $S$.  We adopt $S=2$, which was previously shown to provide a  good quantitative fit to the dielectric properties of SrTiO$_3$.\cite{chapman_modified_2020}

The electrostatic potential $V_{i_z}$ and electric field $E_{i_z}$ that appear in $\hat H_\mathrm{el}$ and $\hat H_\mathrm{TIM}$ are obtained from Gauss' law, applied to the layered geometry and subject to the boundary conditions
\begin{eqnarray} \label{eq:bc}
E(z<-\Delta L) &=& 0, \\
V(z=0) &=& 0,
\end{eqnarray}
where $z$ is the displacement perpendicular to the interface, measured with respect to the interface, and $\Delta L$ is the thickness of the cap layer  (see Fig.~\ref{fig:schematic}).  The first  boundary condition states that the electric field vanishes to the left of the cap layer, while the second defines the electrostatic potential to be zero at the interface. From Eq.~(\ref{eq:bc}), we obtain an expression for the electric displacement inside the FE substrate ($0 \leq z \leq L$),
\begin{equation}
D(z) = \epsilon_\infty E(z) +P(z) = \sum_{i_z < z/a} \sigma^f_{i_z} + \frac{e(n_c + n_\mathrm{2D})}{a^2},
\label{eq:Efield}
\end{equation}
with $\sigma^f_{i_z} = -en_{i_z}/a^2$ the 2D free electron charge density for layer $i_z$. Because the model treats the layers as discrete elements, one may identify
\begin{equation}
P_{i_z} = P \left[ \left( i_z + \frac{1}{2} \right) a \right], \quad E_{i_z} = E \left[ \left( i_z + \frac{1}{2} \right) a \right],
\end{equation}
i.e., $E_{i_z}$ is the electric field between layers $i_z$ and $i_z + 1$. The electrostatic potential can be obtained from Eq.~(\ref{eq:Efield}) by integration. The capacitor charge can then be related to $V_b$ by evaluating the potential difference across the substrate.

\begin{table}
	\begin{tabular}{|c|c||c|c|}
		\hline
		Parameter &  Value & Parameter & Value \\
		\hline
		$t_\|$ & 236 meV & $S$ & 2 \\
		$t_\perp$ & 35 meV & $\epsilon_\infty$ & $5.5\epsilon_0$ \\
		$J_0$ & 7.5~meV & $\epsilon_\mathrm{cap}$ & $25\epsilon_0$ \\
		$J_1$ & 300~meV & $a$ & 3.9~\AA \\
		$\Omega$ & 5.89~meV & $\Delta L$ & 38~\AA\\
		$\mu/a^3$ & 27 $\mu$Ccm$^{-2}$ && \\
		\hline
	\end{tabular}
	\caption{Model parameters.}
	\label{table:params}
\end{table}	

The model is solved iteratively. In the first step, the polarization is obtained from Eq.~(\ref{eq:Bfn}) for a fixed value of the electric displacement and electron density.  In the second step, the eigenstates of the electronic Hamiltonian are obtained for the electrostatic potential corresponding to the electric displacement and polarization obtained in the first step.  From these eigenstates, the free electron density $n_{i_z}$ and corresponding 2D charge density $\sigma^f_{i_z}$ are obtained for each layer using Eq.~(\ref{eq:niz}).  Finally, in the last step, an updated electric displacement may be obtained from Eq.~(\ref{eq:Efield}). To help with convergence of the self-consistent calculations, which can be difficult for this geometry, we perform calculations at $T = 10$~K rather than zero temperature.

The self-consistent polarization and electron density are calculated for the model parameters shown in Table~\ref{table:params}.  Except for $J_0$ and $\mu$, these parameters were obtained by fitting to SrTiO$_3$\cite{chapman_modified_2020} and are presumed to apply to lightly doped SrTiO$_3$ as well.
Relative to their values in SrTiO$_3$\cite{chapman_modified_2020}, we have reduced the parameter $\mu$ slightly to give polarization values similar to those found in Sr$_{1-x}$Ca$_{x}$TiO$_{3-\delta}$ thin films\cite{deLima:2015te}, while we have increased $J_0$ to obtain a FE transition at $T_c = 30$~K, which is comparable to what is measured in insulating  Sr$_{0.98}$Ca$_{0.02}$TiO$_3$ \cite{deLima:2015te}.

From the numerous experiments\cite{deLima:2015te,Rischau:2017vj,engelmayer_ferroelectric_2019} that have shown that a FE-like transition persists in metallic Sr$_{1-x}$Ca$_{x}$TiO$_{3-\delta}$, it has been found that $T_c$ is reduced by electron doping. Clear signatures of a sharp transition were found for electron densities up to $n\sim 10^{-3}$ per unit cell, and for $n\gtrsim 10^{-3}$, the main effect of doping is to broaden the transition.\cite{engelmayer_ferroelectric_2019} Our model does not include this physics. In regions where the 2DEG accumulates, the polarization should therefore be reduced below the value predicted by our calculations. However, $n$ is well below $10^{-3}$ per unit cell throughout most of the substrate, and for this reason we expect that any dependence on electron doping would change the results described below quantitatively but not qualitatively.

\section{Results}
\label{sec:results}

The goal of this section is to establish the effects of metallicity on FE interfaces.  For the  parameters in Table~\ref{table:params}, the bulk polarization is $P_\mathrm{bulk} = 16.4$~$\mu$C$\cdot$cm$^{-2}$ at low $T$; however, this value is reduced significantly for thin films by depolarizing fields, and the polarization vanishes below the critical thickness $L_\mathrm{crit} = 285a$. On the other hand,  electron-doped films screen the depolarizing fields (as we discuss in detail in Sec.~\ref{sec:HighBranch}) and have nonzero lattice polarizations even for films of order a few tens of unit cells thick. We choose $L=200a$ as the substrate thickness for the remainder of this paper; because this is less than the critical thickness of the insulating FE, the average polarization $P_\mathrm{av}$ has a significant dependence on $n_\mathrm{2D}$;  for thicker films with $L\gg L_\mathrm{crit}$, $P_\mathrm{av}$ is close to the bulk value and the dependence on $n_\mathrm{2D}$ is weak.

\begin{figure}[tb]
	\includegraphics[width=\columnwidth]{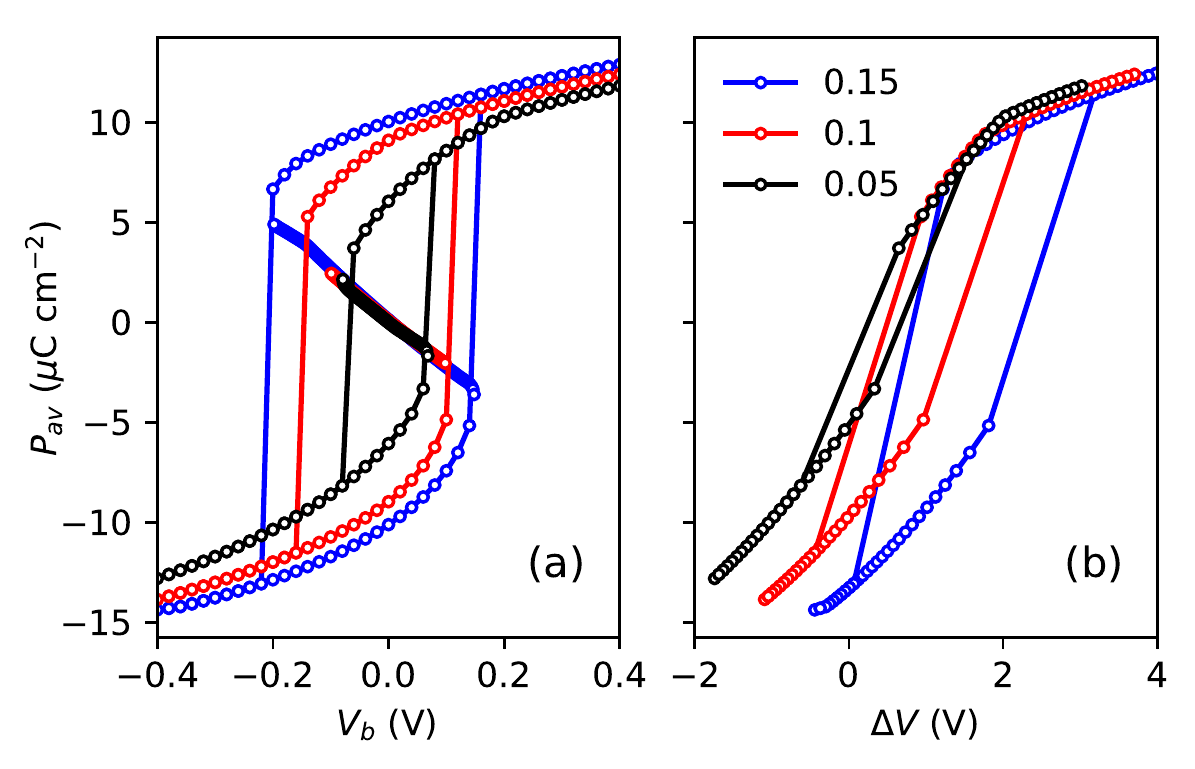}
	\caption{(a) Average polarization as a function of $V_b$, the bias voltage across the substrate, for three different electron densities: $n_\mathrm{2D} = 0.05$, $0.10$, and $0.15$ per 2D unit cell.  Parameters are as in Table~\ref{table:params}, $N_L=200$ layers, and $T=10$~K.  (b)  Average polarization as a function of $\Delta V$, the voltage across the substrate and cap layer.  The low-polarization branch is not shown in (b). }
	\label{fig:hysteresis}
\end{figure}

We begin with the simplest point of comparison, namely the average lattice polarization of the FE film,
\begin{equation}
P_\mathrm{av} = \frac{1}{L} \int_0^L dz P(z) = \frac{1}{N_L} \sum_{i_z = 0}^{N_L-1} P_{i_z}.
\end{equation}
Figure~\ref{fig:hysteresis} establishes that in  the metallic films $P_\mathrm{av}$  is both switchable and hysteretic as a function of bias voltage.  Figure~\ref{fig:hysteresis}(a) shows the average polarization as a function of $V_b$ for three different values of $n_\mathrm{2D}$.   For each of these, there are two high-polarization branches with $|P_\mathrm{av}|\sim 5$--15~$\mu$C$\cdot$cm$^{-2}$, and a low-polarization branch with $|P_\mathrm{av}|\approx 0$ when $V_b=0$.  In all cases, the high-polarization branches exhibit hysteresis.  In Fig.~\ref{fig:hysteresis}(b), we demonstrate that the hysteresis is also present when the polarization is plotted as a function of $\Delta V$, although it is obscured by the fact that $\sim 90\%$ of the voltage drop is across the cap layer, and only $\sim 10\%$ is across the substrate. 

The 2DEG has two clear effects on the hysteresis loops.  First, as mentioned above, $|P_\mathrm{av}|$ increases with increasing $n_\mathrm{2D}$ because depolarizing fields are increasingly screened; second, the coercive bias voltage at which the polarization switches sign increases with $n_\mathrm{2D}$.  The shift in coercive voltage is due both to changes in $|P_\mathrm{av}|$ and to the screening of external fields by the 2DEG.  Because the interface  breaks the mirror symmetry of the substrate, this screening affects the upper and lower branches of the hysteresis curves differently, so that the overall hysteresis pattern is asymmetric.  These points are discussed in detail in Sec.~\ref{sec:HighBranch} and Sec.~\ref{sec:discussion}.

There is, in addition, a low-polarization branch with $P_\mathrm{av}=0$ at $V_b=0$.   This branch has a negative slope as a function of $V_b$, which suggests that the FE has a negative dielectric susceptibility. In insulating FEs, the negative slope is an indication that states on that branch are unstable; however, it has  been shown recently that dielectric phases with negative susceptibility may be stabilized in heterostructures provided the overall capacitance of the heterostructure is positive.\cite{appleby:2014,zubko:2016,hoffmann:2019}  These points are explored further in Sec.~\ref{sec:LowBranch}, where we show that the low-polarization branch corresponds to a phase-separated state with two FE domains separated by a head-to-head domain wall.  

\subsection{High-Polarization Branches}
\label{sec:HighBranch}

\begin{figure*}[tb]
	\includegraphics[width=\textwidth]{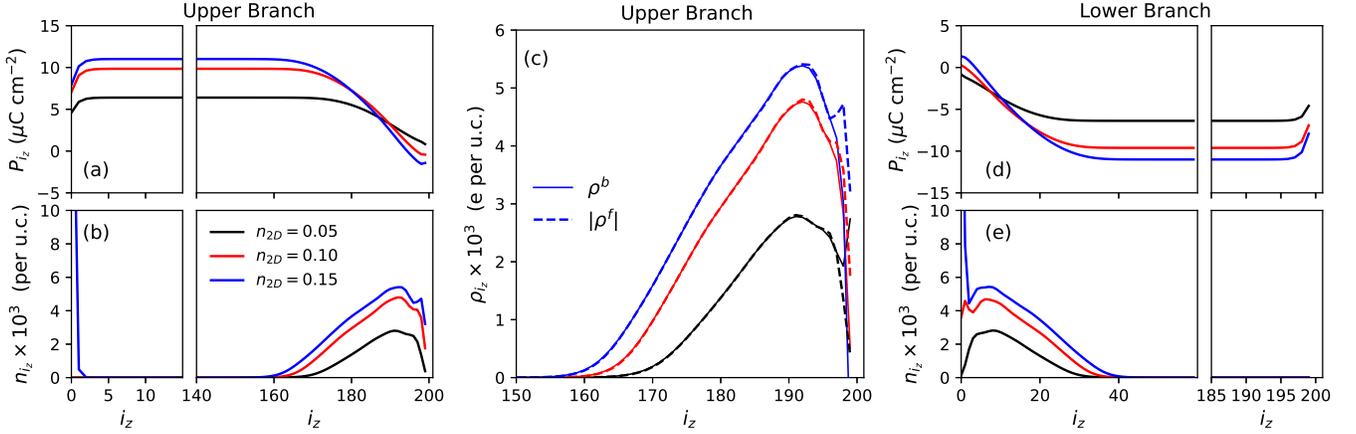}
	\caption{Polarization and electron density profiles in the short-circuit configuration, $V_b=0$, for three different values of the electron density.   (a) Polarization and (b) electron density profiles are shown as a function of layer index for the upper branch of Fig.~\ref{fig:hysteresis}.  (c) The free and bound charge densities per layer, $\rho^f$ and $\rho^b$, respectively, are shown for the final 50 layers of the ferroelectric substrate.  The surface polarization charge ${\bf P}\cdot {\bf \hat n}$ is not shown. (d) Polarization and (e) electron density profiles are shown for the lower branch of Fig.~\ref{fig:hysteresis}. }
	\label{fig:upbranch_profs_Vb0}
\end{figure*}

In this section, we explore the structure of both the polarization and the 2DEG for states belonging to the high-polarization branches of Fig.~\ref{fig:hysteresis}.  
To begin, we show in Fig.~\ref{fig:upbranch_profs_Vb0} results for the polarization and electron density  at zero bias voltage ($V_b = 0$). Figure~\ref{fig:upbranch_profs_Vb0}(a) shows upper branch polarization profiles
for $n_\mathrm{2D} = 0.05$, 0.10, and 0.15 electrons per 2D unit cell; the corresponding electron densities are shown in Fig.~\ref{fig:upbranch_profs_Vb0}(b).  These figures focus on regions near both the interface (at $i_z=0$) and the back of the substrate ($i_z = 199$).

Figure~\ref{fig:upbranch_profs_Vb0}(a) shows that, except near the substrate surfaces, the polarization $P_{i_z}$ obtains a uniform value $P^0$ that depends on the total electron density.  Near the interface, $P_{i_z}$ varies abruptly over a few layers before saturating at $P^0$; at the back wall, $P_{i_z}$ changes smoothly over a much longer length scale. From Fig.~\ref{fig:upbranch_profs_Vb0}(b), the region where the polarization varies smoothly contains most or all of the 2DEG.  The distinction between the two regions is thus whether or not the depolarizing fields due to polarization gradients are screened by the 2DEG.

The screening arises because the charge density associated with the 2DEG compensates the bound charge density,
\begin{equation}
\rho^b(z) = -\frac{d}{dz}P(z),
\label{eq:rhob}
\end{equation}
associated with polarization gradients.  The two charge densities are compared in Fig.~\ref{fig:upbranch_profs_Vb0}(c).   Remarkably, $\rho^b_{i_z} \approx -\rho^f_{i_z}$ everywhere except for the final few layers nearest the back wall of the FE substrate.  The 2DEG and lattice polarization compensate each other, such that the net charge density vanishes away from the surface. This compensation was noticed previously in simulations of non-FE LaAlO$_3$/SrTiO$_3$ interfaces and is a generic feature of materials that have large FE correlation lengths relative to the Fermi wavelength.\cite{Atkinson:2017jt}  

We note that, for the largest electron density shown in Fig.~\ref{fig:upbranch_profs_Vb0} ($n_\mathrm{2D}=0.15$), there is a component of the 2DEG that resides at the interface. As we show below, this is a spillover effect that occurs when the electron density exceeds what is needed to screen the polarization gradients.  Here, we simply note this excess charge cannot screen the polarization gradients at the interface, since these have a \textit{negative} charge ($\rho^b < 0$).

In regions where depolarizing fields are screened, the characteristic length scale for spatial gradients of the polarization is the Landau-Ginzburg-Devonshire (LGD) correlation length, $\xi$.   We can obtain an estimate for $\xi$ by using the relationship between the transverse Ising model and the simplest LGD functional,\cite{chapman_modified_2020}
\begin{equation}
{\cal F} = \int d^3r\left [ \frac 12 AP^2 + \frac 14 BP^4 + \frac 12 C |\nabla P|^2 \right ]. 
\end{equation} 
For the parameters given in Table~\ref{table:params}, we obtain the $T=0$ result
\begin{equation}
\xi = \sqrt{\left| \frac{C}{2A} \right |} = \sqrt{\frac{J_1 a^2}{2(J_0 - \Omega)}} \approx 10a.
\end{equation}  
This is the longer of the two length scales identified in Fig.~\ref{fig:upbranch_profs_Vb0}.  

Within a few monolayers of each surface, where the free electron density is small, depolarizing  fields are unscreened and the LGD energy is
\begin{equation}
\tilde{\cal F} = {\cal F} - \frac 12 \int d^3r E_d P,
\end{equation}
where $E_d = -\frac{1}{\epsilon_\infty} P$ is the depolarizing field.  The depolarization term in $\tilde{\cal F}$ is therefore quadratic in $P$, so that the renormalized correlation length is\cite{kretschmer79}
\begin{equation}
\xi_0 = \sqrt{\frac{-C}{2(\epsilon_\infty^{-1}+A)}} 
=\sqrt{\frac{-J_1a^2}{2(\Omega-J_0 + \frac{\mu^2}{a^{3}S^2\epsilon_\infty})}} \approx a.
\end{equation}  
This is the shorter of the two length scales identified in Fig.~\ref{fig:upbranch_profs_Vb0}.

\begin{figure}[tb]
	\includegraphics[width=\columnwidth]{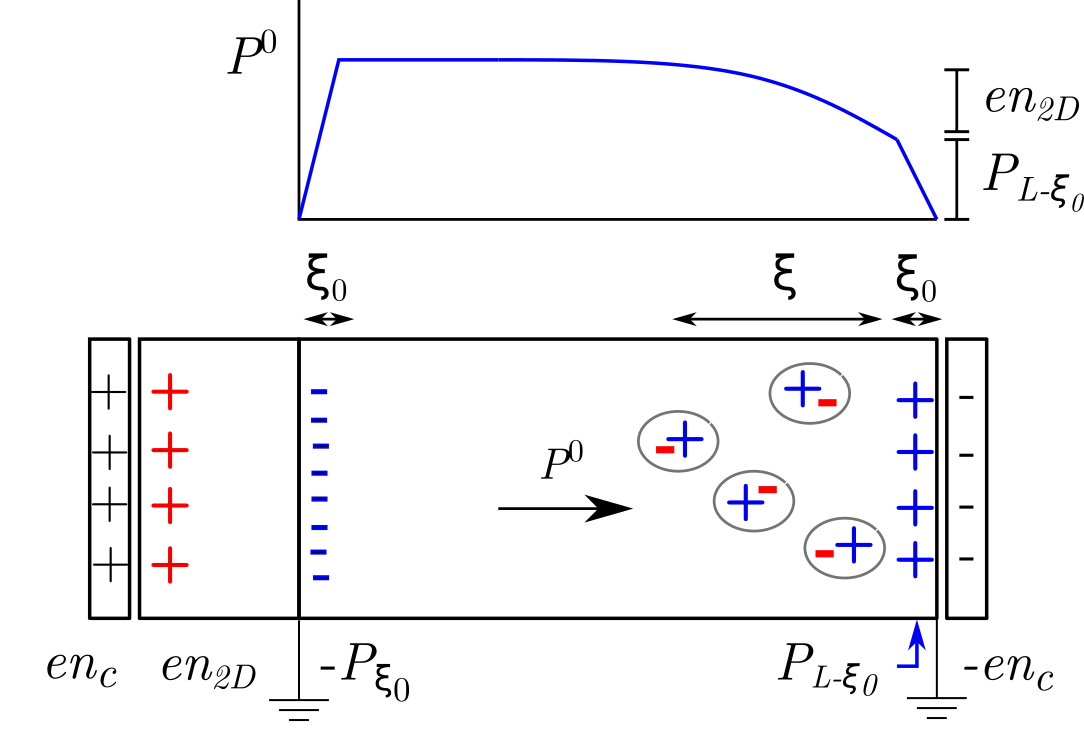}
	\caption{Cartoon showing the charge distribution and polarization profile for the high-polarization branch in the short-circuit configuration and low electron densities, $en_\mathrm{2D} < P^0$.   The two capacitor plates have charge densities $\pm en_c$; there is a charge density $en_\mathrm{2D}$ on the surface of the cap layer, represented by the red plus signs; and the substrate has surface charge densities $-P_{\xi_0}$ and $P_{L-\xi_0}$ due to the lattice polarizations, represented by the blue minus and plus signs at the surfaces, respectively.  Polarization gradients within the substrate are compensated by the 2DEG (shown as the circled charges in the figure).  Overall, the system is charge-neutral, and the charges are arranged such that the electric field in the substrate would be zero in the absence of quantum effects. }
	\label{fig:cartoon}
\end{figure}

\begin{figure*}[tb]
	\includegraphics[width=\textwidth]{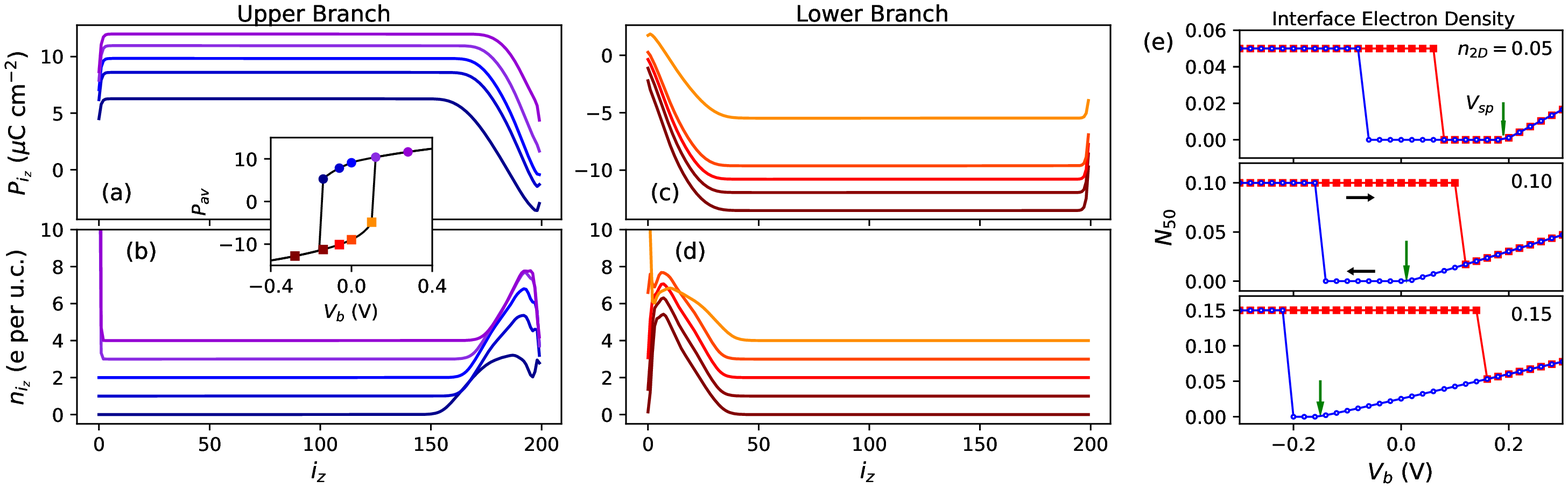}
	\caption{Effects of bias voltage for states on the (a, b) upper and (c, d) lower branches of the hysteresis curve for $n_\mathrm{2D}=0.10$.  Profiles are shown for (a, c) the polarization and (b, d) the electron density at discrete values of the bias voltage, $V_b$. These values are indicated on the hysteresis curve plotted in the inset. Note that symbol colors in the inset match the colors in (a-d). The electron density profiles (b, d) have been offset vertically for clarity, but the polarization profiles (a, c) are not offset. (e)  The integrated electron density within 50 layers of the interface is shown as a function of $V_b$ for $n_\mathrm{2D} = 0.05$, 0.10, and 0.15.  The  voltage $V_\mathrm{sp}$, above which electrons spill over to the interface in the positive polarization configuration, is indicated by a vertical arrow in each panel.}
	\label{fig:Vb}
\end{figure*}

We thus arrive at the cartoon in Fig.~\ref{fig:cartoon}, which is meant to emphasize the roles of the two length scales and of the charge compensation by the 2DEG.  In an insulating FE, the polarization would be $P^0$ everywhere, except within $\sim O(\xi_0)$ of the surfaces;  this would generate surface charge densities ${\bf P}\cdot {\bf \hat n}$ equal to $-P^0$ and $P^0$ at the interface and back wall, respectively; these are distributed over depths $\sim O(\xi_0)$, and are the source of the depolarizing field.  In the metallic FE, the surface charge density remains $-P^0$ at the interface (although the value of $P^0$ may be different), and an equal and opposite charge density appears at the back wall; however, this is broken into two components.  The first component is compensated by the 2DEG and spreads into the substrate a distance $\sim O(\xi)$.  The second component remains attached to the surface, and is
\begin{equation}
{\bf P}\cdot {\bf \hat n} = P_{L-\xi_0} = P^0 - e n_\mathrm{2D}.
\end{equation}
The depolarizing electric field due to these surface charges is 
\begin{equation}
\epsilon_\infty E_d = -\frac{P_{\xi_0} +P_{L-\xi_0}}{2} 
= -\left( P^0 - \frac 12 en_\mathrm{2D}\right ).
\end{equation}
Depolarizing fields are thus screened by the 2DEG.

These considerations also suggest that when $e n_\mathrm{2D} > P^0$, the polarization will change sign at the back wall of the substrate, and indeed this is what is seen in Fig.~\ref{fig:upbranch_profs_Vb0}(a): the polarization remains positive when $e n_\mathrm{2D} < P^0$ ($n_\mathrm{2D} = 0.05$) but is inverted at the the back wall when $e n_\mathrm{2D} >  P^0$ ($n_\mathrm{2D} = 0.15$).  However, this effect is small because most of the excess electron density migrates to the interface as a result of attraction to the positive charges on the surface of the cap layer.  In other words, most of the excess electron density, which is not required to screen depolarizing fields, screens the external electric fields.

So far, the discussion has included only states belonging to the upper branches of the hysteresis curves.  Figures~\ref{fig:upbranch_profs_Vb0}(d) and \ref{fig:upbranch_profs_Vb0}(e) show polarization and electron density profiles for the lower branches (again, at $V_b=0$).  When $en_\mathrm{2D} < P^0$ ($n_\mathrm{2D} = 0.05$), the polarization and electron densities are mirror images of those shown in Figs.~\ref{fig:upbranch_profs_Vb0}(a) and \ref{fig:upbranch_profs_Vb0}(b) for positive polarizations.  However, this symmetry is broken when $en_\mathrm{2D} > P^0$ ($n_\mathrm{2D} = 0.15$), because the excess electrons (beyond what are needed to screen polarization gradients) remain at the interface [Fig.~\ref{fig:upbranch_profs_Vb0}(e)].  When $en_\mathrm{2D} > P^0$, then, the interface is metallic for both polarization states at $V_b = 0$.  This has important consequences: the existence of a switchable lattice polarization does not guarantee that the 2DEG is also switchable.

Figure~\ref{fig:Vb} shows polarization and electron densities as a function of the substrate bias $V_b$.  Figs.~\ref{fig:Vb}(a)-(d) focus on the case $n_\mathrm{2D} = 0.10$ and show results for both upper and lower branches of the hysteresis curve shown in the figure inset.  In Figs.~\ref{fig:Vb}(a) and (b), the middle curve corresponds to $V_b = 0$. By definition, a positive $V_b$ corresponds to a potential that is higher at the interface than at the back wall; this increases the average polarization and confines the 2DEG closer to the back wall of the FE substrate.  However, for large enough $V_b$, a fraction of the 2DEG spills over to the interface, where it partially screens the external electric field.  The onset voltage $V_\mathrm{sp}$ at which this spillover happens is marked by a kink in the upper branch of the hysteresis curve, most easily seen in Fig.~\ref{fig:hysteresis}(b).  Clearly, the screening is incomplete, as $P_\mathrm{av}$ continues to increase with increasing $V_b$ beyond $V_\mathrm{sp}$; however, the rate of increase is reduced compared to voltages smaller than the spillover voltage.

Figures~\ref{fig:Vb}(c) and (d) show polarization and electron density profiles along the lower branch of the hysteresis curve.  Here, the 2DEG resides entirely at the interface for all values of $V_b$.  Sufficiently close to the coercive field, the polarization drops below $en_\mathrm{2D}$ and the 2DEG separates into a component that is tightly bound to the interface by external fields, and a loosely bound component that screens the polarization gradients.  The charge profiles are therefore asymmetric under reversal of $V_b$.

To characterize the behavior of the 2DEG as a function of $V_b$, we plot in Fig.~\ref{fig:Vb}(e) the total electron density $N_\mathrm{50}$ contained within 50 layers of the interface.  The figure shows clearly the voltage $V_\mathrm{sp}$ above which the interface becomes metallic along the positive polarization branch (the lower curve in all panels).  Importantly, $V_\mathrm{sp}$ is a strong function of electron density and is largest for small $n_\mathrm{2D}$.

\begin{figure}[tb]
	\includegraphics[width=\columnwidth]{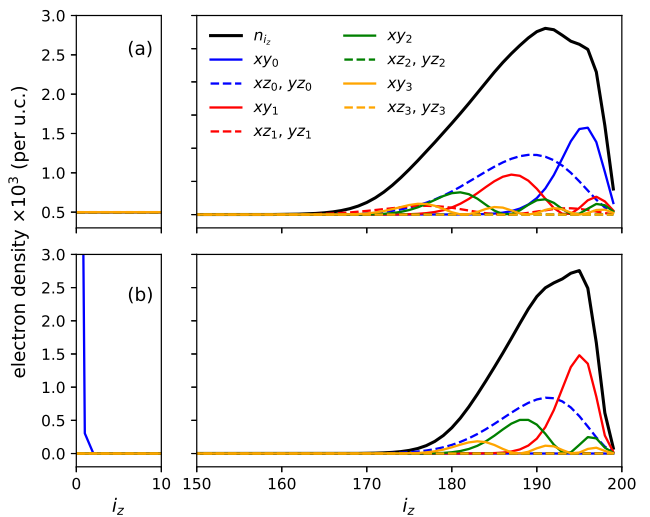}
	\caption{Contributions of individual bands to the electron density for the high-polarization state with $n_\mathrm{2D}=0.05$ and (a) $V_b=0$ or (b) $V_b = 0.3$~V.  The figure shows the total electron density $n_{i_z}$ and the weighted contributions of the bands making the largest contributions to the electron density. The different bands are denoted by $\alpha_n$, where $\alpha = xy$, $xz$, $yz$ are the orbital symmetries and $n$ is the band index.}
	\label{fig:orbitals}
\end{figure}

The orbital characters of the bands making the largest contributions to the 2DEG are shown for states on the upper branch of the hysteresis curve for $n_\mathrm{2D} = 0.05$ at $V_b = 0$ [Fig.~\ref{fig:orbitals}(a)] and $V_b = 0.3$~V [Fig.~\ref{fig:orbitals}(b)].  The spatial profiles shown in the figure are obtained from the eigenfunctions of the electronic  Hamiltonian, Eq.~(\ref{eq:Hel}), and the band filling defined following Eq.~(\ref{eq:niz}). Each curve in Fig.~\ref{fig:orbitals} (besides the $n_{i_z}$ curve) thus corresponds to a single term in the sum in Eq.~(\ref{eq:niz}). Almost all of the charge density in the figure is contained in the four lowest-energy $xy$ and two lowest-energy $xz$ and $yz$ bands.  The occupation differences between $xy$ and $xz$, $yz$ bands reflects an orbital selectivity coming from surface effects, namely that the $xy$ bands are heavy along the $z$-direction and are more easily confined by a potential than the lighter $xz$ and $yz$ bands.   This effect is well-known in LaAlO$_3$/SrTiO$_3$ interfaces, and is expected to be pronounced at ideal polar interfaces.\cite{Gariglio:2015jx}  Here, it is comparatively weak because the confining potential well due to the polarization discontinuity at the back wall is small.  In contrast, there is a strong confining potential at the interface that becomes relevant when $V_b > V_\mathrm{sp}$, as in Fig.~\ref{fig:orbitals}(b).  In this case, there is a low energy band with $xy$ orbital symmetry that is confined to the first layer.

\subsection{Low-polarization branch}
\label{sec:LowBranch}

\begin{figure}[tb]
	\includegraphics[width=\columnwidth]{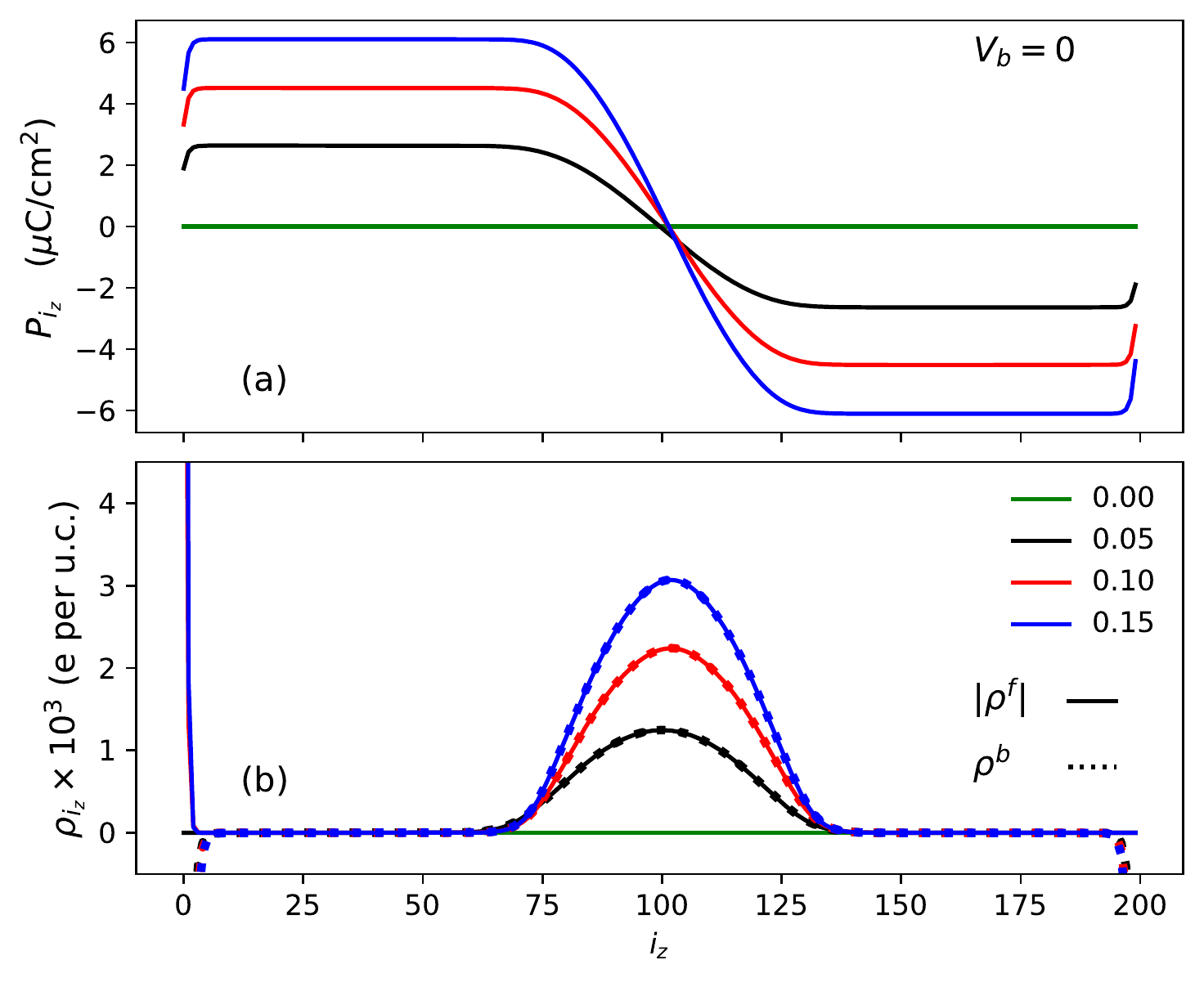}
	\caption{Profiles of states belonging to the low-polarization branch at $V_b=0$.  (a) Polarization and (b) charge density profiles are shown for $n_\mathrm{2D} = 0.0$, 0.05, 0.10, and 0.15 electrons per 2D unit cell.   In (b), solid (dotted) curves represent $|\rho^f_{i_z}|$ ($\rho^b_{i_z}$).  When $n_\mathrm{2D} = 0.05$, the 2DEG is bound to the domain wall; when $n_\mathrm{2D} = 0.10$ and 0.15,  the 2DEG is shared between the domain wall and the interface. }
	\label{fig:Pave0}
\end{figure}

Figure~\ref{fig:Pave0} shows a comparison of the polarization [Fig.~\ref{fig:Pave0}(a)] and free and bound charge density profiles [Fig.~\ref{fig:Pave0}(b)] for insulating ($n_\mathrm{2D} = 0.0$) and metallic films in the low-polarization state, in the short-circuit geometry.  In all cases, the average polarization vanishes.  Such solutions may always be found in the short-circuit geometry, and they are energetically stable (metastable) when $L$ is below (above) the critical film thickness.  The striking feature of this figure is that the polarization profiles are completely different for the two kinds of film:  For the insulating FE, $P_{i_z}$ is nearly uniform and is nearly zero everywhere;  for the metallic FE, $P_{i_z}$ is roughly 50\% of its value on the high-polarization branches, and $P_\mathrm{av}$ vanishes because it is energetically favorable for a head-to-head domain wall to form at the center of the film.   

Head-to-head domain walls were found for all film thicknesses that we studied. In the smallest systems, where the sample thickness is comparable to the correlation length $\xi$, the domain wall spans the entire width of the sample:  that is, the polarization is positive (negative) near the front (back) surface and interpolates approximately linearly between the two surfaces.  We found no solutions for $T<T_c$ in which the local polarization $P_{i_z}$ vanished when $n_\mathrm{2D}$ is nonzero.

Figure~\ref{fig:Pave0}(b) shows that most or all of the 2DEG is bound to the domain wall, depending on the value of $n_\mathrm{2D}$.  At low electron densities ($n_\mathrm{2D}=0.05$), the 2DEG is confined entirely to the domain wall; at higher densities ($n_\mathrm{2D} =0.10$, 0.15), a small fraction of the 2DEG spills over and is bound to the interface.

Furthermore, Fig.~\ref{fig:Pave0}(b) shows that the domain walls are close to being electrically neutral.  Except at the surfaces, $\rho^f_{i_z} \approx -\rho^b_{i_z}$, similar to the high-polarization case  [Fig.~\ref{fig:upbranch_profs_Vb0}(c)].  As we show below, this cancellation is not perfect and there is a small net residual charge that determines the response of the  domain wall to an applied electric field; however, to a first approximation, one may think of them as neutral. Because of this, the electrostatic cost to form a  domain wall (nearly) vanishes, and the overall energetic cost of formation is dramatically reduced.  Furthermore, because electric fields are screened by the 2DEG, the width of the domain wall is set by the correlation length $\xi$, rather than the much shorter length $\xi_0$ that is relevant for insulating FEs.

The magnitude of the polarization in each of the domains is reduced by a factor of $\sim 3$--6 from the bulk polarization, and to a first approximation is set by the requirement of domain-wall neutrality.   The net 2D charge across the  domain wall is 
\begin{equation}
\sigma_\mathrm{domain} = \int \rho^b(z) dz -en_\mathrm{DW}
= 2P^0 -en_\mathrm{DW},
\end{equation}
where $P^0$ ($-P^0$) is the value of the polarization to the left (right) of the domain wall in Fig.~\ref{fig:Pave0}(a) and $n_\mathrm{DW}$ is the two-dimensional electron density in the  domain wall.  The condition of neutrality requires that 
\begin{equation}
P^0 = \frac 12 e n_\mathrm{DW} \leq \frac 12 e n_\mathrm{2D},
\label{eq:dw_neutrality}
\end{equation}
and we see in Fig.~\ref{fig:Pave0}(a) that $P^0$ indeed grows with increasing $n_\mathrm{2D}$.
At low electron density ($n_\mathrm{2D} = 0.05$), $n_\mathrm{DW} = n_\mathrm{2D}$; at higher electron densities, where some of the 2DEG spills over to the interface, $n_\mathrm{DW} < n_\mathrm{2D}$.

\begin{figure}[tb]
	\includegraphics[width=\columnwidth]{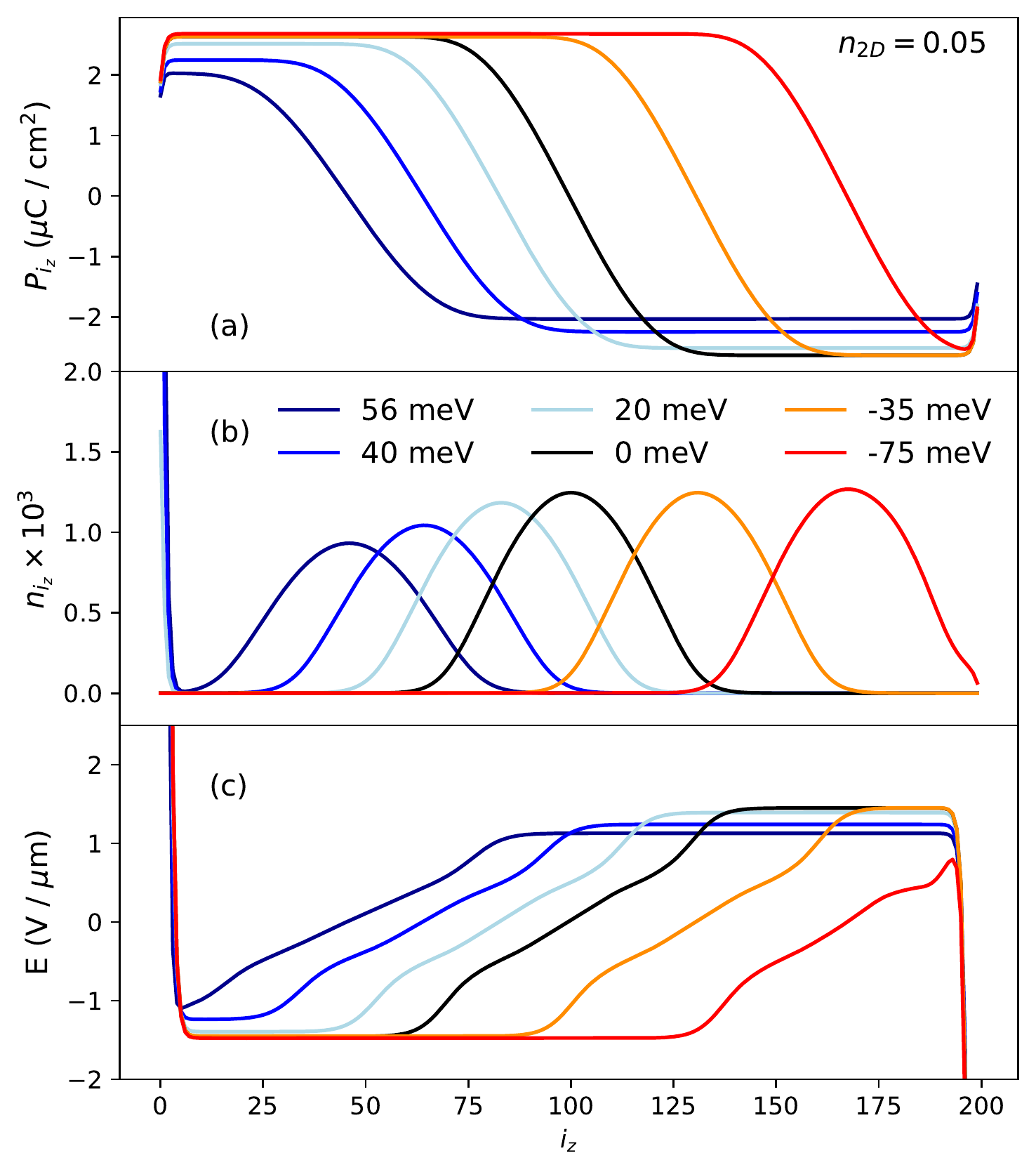}
	\caption{Profiles of states belonging to the low-polarization branch with $n_\mathrm{2D}=0.05$.  (a) Polarization, (b) electron density, and (c) electric field profiles are shown for $V_b = -75$, \ldots, 56~meV.  In (c), the electric field at the surfaces is approximately two orders of magnitude larger than inside the film. }
	\label{fig:Pave1}
\end{figure}

For the low-polarization branches of the hysteresis curves (Fig.~\ref{fig:hysteresis}), the average polarization decreases (increases) with increasing (decreasing) $V_b$,  which corresponds to a negative effective dielectric susceptibility for the FE substrate.
To investigate this dielectric response, we show in Fig.~\ref{fig:Pave1} the responses of the electron density, polarization, and electric field profiles 
to nonzero bias voltages for the case $n_\mathrm{2D} = 0.05$. The key result of this figure is that the voltage-dependence of $P_\mathrm{av}$ in the low-polarization branch occurs because the domain wall and 2DEG move in opposition to the applied electric field:  Under a positive bias voltage, they  shift toward the interface, and under a negative bias voltage they shift toward the back wall [Figs.~\ref{fig:Pave1}(a) and (b)].
This is quite different from the high-polarization branches, where the {\em magnitude} of the polarization in the FE substrate is a strong function of $V_b$ (Fig.~\ref{fig:Vb}).

The anomalous domain wall motion is governed by the internal electric fields, shown in Fig.~\ref{fig:Pave1}(c).  The external electric fields are strongly screened by the FE substrate and fall by two orders of magnitude within a few lattice constants of the surfaces.  The residual fields within the substrate are determined by the domain wall's net charge, which is seen to be positive since $dE/dz > 0$ in the domain wall region.

A negative value of $V_b$ means that the potential at the interface is less than that at the back wall of the substrate, or equivalently that $\int_0^L E dz < 0$.  This is achieved in Fig.~\ref{fig:Pave1} by moving the domain wall to the right; paradoxically, this motion increases $P_\mathrm{av}$ and leads to the negative dielectric susceptibility, namely $dP_\mathrm{av}/dV_b < 0$.

Similar considerations apply when $V_b >0$; however, there is an additional effect, namely 
there is a voltage-dependent spillover of charge to the interface from the domain wall. Similar to the high-polarization branches, the electron spillover occurs above a voltage $V_\mathrm{sp}$.  The value of $V_\mathrm{sp}$ is different from the high-polarization branches, however, and for $n_\mathrm{2D}=0.05$ it lies (by coincidence) near 0~meV. This crossover is clear in Figs.~\ref{fig:Pave1}(a) and (b): For $V_b < V_\mathrm{sp}$, $n_\mathrm{DW} = n_\mathrm{2D}$ and $P^0$ is independent of bias voltage; for $V_b > V_\mathrm{sp}$, both $n_\mathrm{DW}$ and $P^0$ are decreasing functions of $V_b$.

Remarkably, the electrons that spill over to the interface do not contribute to the internal fields in the substrate: The electric field on the left of the domain wall is equal and opposite to that on the right, which indicates that the residual field in the substrate is due entirely to the domain wall. As we shall see below, this is because the growth of the interfacial 2DEG is compensated by a transfer of charge between the capacitor plates enclosing the heterostructure  (Fig.~\ref{fig:schematic}).

\begin{figure}[tb]
	\includegraphics[width=\columnwidth]{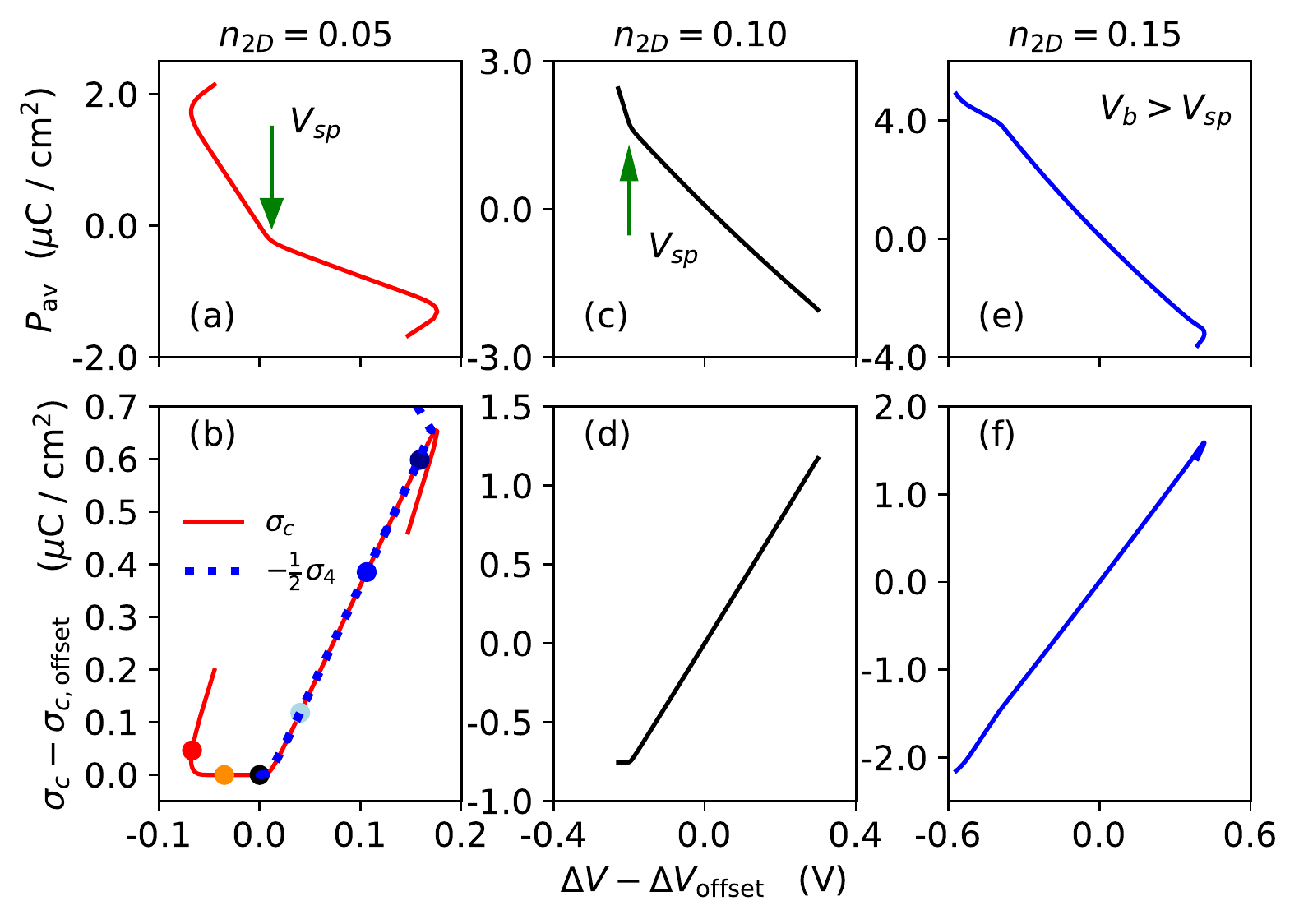}
	\caption{Average polarization (a, c, e) and surface charge density (b, d, f) on the capacitor plates as a function of voltage bias across the capacitor for (a, b)  $n_\mathrm{2D}=0.05$, (c, d) $n_\mathrm{2D}=0.10$, and (e, f) $n_\mathrm{2D}=0.15$ electrons per 2D unit cell. Spillover voltages are indicated in (a) and (c), while $V_b > V_\mathrm{sp}$ over the entire voltage range in (e).  In (b), the interfacial electron charge density $\sigma_4$ (summed over the first four layers) multiplied by $-\frac 12$ is shown for $V_b > V_\mathrm{sp}$. Both the voltage bias and charge density are shifted by offsets such that the point $(0,0)$ corresponds to the short-circuit case, with $V_b=0$.}
	\label{fig:Pave2}
\end{figure}

We plot in Fig.~\ref{fig:Pave2} the charge $\sigma_c = e n_c$ on the top capacitor plate  as a function of the voltage $\Delta V$ across the capacitor, along with $P_\mathrm{av}$. For clarity, both $\sigma_c$ and $\Delta V$ have been shifted by an offset in such a way that the point $(0, 0)$ on the figure corresponds to the short-circuit case, $V_b=0$. The slope $d\sigma_c/d\Delta V$ represents the differential capacitance. Three distinct behaviors are observed in this figure:  First, in all cases there are voltage ranges over which $d\sigma_c/d\Delta V > 0$, indicating that the device has a positive capacitance; second, in Figs.~\ref{fig:Pave2}(b) and (d) there are voltage ranges over which $\sigma_c$ is constant, such that the capacitance vanishes; and third, in Figs.~\ref{fig:Pave2}(b) and (f) the $\sigma_c$ curve ``folds over'' at the ends so that it is a multi-valued function of $\Delta V$.  From the plots of $P_\mathrm{av}$ in Figs.~\ref{fig:Pave2}(a) and (e), we see that the ``folding over'' marks the transition from the low-polarization to high-polarization branches, and we will not discuss it further.  We focus instead on understanding the distinction between regimes in which $\sigma_c$ is constant and those over which it grows linearly with $\Delta V$.

Regimes of constant $\sigma_c$ correspond to ranges of $V_b < V_\mathrm{sp}$ for which there is no electron spillover, such that the 2DEG is entirely confined to the domain wall. In this regime, changes in $\Delta V$ are due entirely to the motion of the domain wall, and not to any transfer of charge between the capacitor plates.  

In contrast, the regimes in which $\sigma_c$ grows linearly with $\Delta V$ have $V_b > V_\mathrm{sp}$ and are characterized by a voltage-dependent transfer of electrons to the interface.  To quantify this transfer, the integrated electron charge density over the first four layers of the FE substrate,
\begin{equation}
\sigma_4 = -\frac{e}{a^2} \sum_{i_z=0}^3 n_{i_z},
\end{equation}
is plotted in Fig.~\ref{fig:Pave2}(b). For the charge on the capacitor to cancel the electric field inside the substrate due to $\sigma_4$, we require $2\sigma_c + \sigma_4 = 0$, where the factor of 2 is because there are two capacitor plates.  In Fig.~\ref{fig:Pave2}(b), we see that indeed $\sigma_c = -\frac 12 \sigma_4$ up to where the curve folds over to the high-polarization branch. Thus, the electric field  due to the interfacial 2DEG is compensated by the charge transfer between the capacitor plates.

We thus arrive at the following picture of the low-polarization branch:  The voltage dependence of $P_\mathrm{av}$, and in particular the negative slope ($dP_\mathrm{av}/d\Delta V < 0$), can be attributed to the motion of the head-to-head domain wall that forms near the center of the film. While the negative slope suggests that the FE substrate has a negative capacitance, the actual capacitance of the device shown in Fig.~\ref{fig:schematic} is always either zero or positive.  Where the capacitance vanishes, the 2DEG in the substrate is entirely bound to the domain wall; where the capacitance is positive, a voltage-dependent component of the 2DEG spills over to the interface. The induced charge on the capacitor plates is equal and opposite to that of the interfacial component of the 2DEG, such that their combined contribution to the electric field in the substrate vanishes. The interfacial charge therefore only affects the domain wall motion indirectly, via the magnitude $P^0$ of the polarization in the film, which depends on the electron density $n_\mathrm{DW}$ in the domain wall.

\section{Discussion and Conclusions}
\label{sec:discussion}

We have explored a model for an interface between an insulating cap layer and a FE substrate, in which it is presumed that doping of the substrate occurs through charge transfer from the surface of the cap layer.  While our model was inspired by a hypothetical LaAlO$_3$/Sr$_{1-x}$M$_x$TiO$_3$ interface, with $M=\mathrm{Ba},\, \mathrm{Ca}$, the primary goal of this work is to identify general physical properties that emerge from electrostatic considerations.   In this way, our work is complementary to earlier works espousing chemical design principles to optimize ferroelectricity in metallic compounds.\cite{puggioni:2014,benedek_ferroelectric_2016}

We found that the average polarization $P_\mathrm{av}$ has a very conventional-looking ``S''-shaped dependence on the bias voltage $V_b$ across the substrate, which we describe as consisting of two high-polarization branches and a single low-polarization branch with a negative effective dielectric susceptibility. This finding negates reasonable concerns that the electron gas might screen external fields and suppress control of the polarization state.

We have identified the mechanism by which switchability is enabled. In simple terms, we might expect the electron gas to perform two tasks:  the screening of internal depolarizing fields, and the screening of external bias fields that are necessary to switch the polarization state. We found that at low electron densities and weak external fields, the electron gas binds to gradients in the lattice polarization to form a neutral compensated state. The electron gas is thus prevented from screening external fields. For both the high- and low-polarization branches, the primary role of the 2DEG is to screen depolarizing fields.

When either the electron density or bias voltage is larger than a spillover threshold, the electron gas develops a component that is not compensated by the bound charge density.  For the geometry used in this work, this component forms a tightly bound state against the interface.  The interfacial electrons occupy a single $d_{xy}$-derived band that extends only a few unit cells into the substrate.  The filling of this band depends on bias voltage, and this state therefore partially screens external fields.  We described this as a ``spillover'' effect.

One important consequence of the electron spillover is that the hysteresis curve for the 2DEG is not the same as for the polarization.  In particular, if the desire is to have a switchable and hysteretic conductivity in the interface region, then performance of the device will be degraded by any residual electron gas at the interface when the polarization switches. Our results suggest that this may be avoided by reducing $n_\mathrm{2D}$ so that it is less than the polarization $P^0$ within the FE substrate.  However, this should be balanced against the desirable effect that, until spillover occurs, an increase in $n_\mathrm{2D}$ more efficiently screens depolarizing fields.  We also note that, for the geometry considered in this work, spillover does not affect the back wall of the FE substrate and that devices based on the conductance at the back wall should be free of unwanted residual metallicity.

The negative slope of the $P_\mathrm{av}$--$V_b$ plots for the low-polarization branch in Fig.~\ref{fig:hysteresis}(a) is reminiscent of the unstable part of the polarization curve predicted by Landau-Devonshire theory.  In recent years, this branch has taken on importance as a way to reduce the power consumption of field-effect transistors.  Key to this is that a bilayer comprising a dielectric and a FE can have a capacitance that is larger than that of the dielectric alone.\cite{appleby:2014,zubko:2016,hoffmann:2019,hoffmann:2021}
We emphasize that the situation presented in this work is different: In the dielectric/FE bilayers, the low-polarization branch is stabilized by the dielectric layer; in the current work, the low-polarization branch is stabilized by the formation of a domain wall in the FE substrate.  This distinction is important: Rather than being enhanced, the capacitance of the device pictured in Fig.~\ref{fig:schematic} may even vanish for certain voltage and electron density ranges.

In summary, we have found that electron-doped FE interfaces, in which the doping occurs via a charge transfer from an insulating cap layer, are fundamentally influenced by the tendency for the conduction electrons to form a neutral compensated state with bound charges resulting from polarization gradients.  This significantly reduces
the ability of the conduction electrons to screen external fields, and permits control of both the polarization direction and electron distribution profiles with an applied voltage bias. While we specifically modelled interfaces based on SrTiO$_3$, we believe that this mechanism should apply broadly to other FE substrates.

\begin{acknowledgments}
	We acknowledge support by the Natural Sciences and Engineering Research Council (NSERC) of Canada.  
\end{acknowledgments}

\bibliography{FE_Interface}

\end{document}